\documentclass[preprint,aps,showpacs]{revtex4}
\usepackage{amssymb} \usepackage{graphicx}

\begin{document}
\title{Killing-Yano Forms of a Class of Spherically Symmetric
 Space-Times I: A Unified Generation of Killing Vector Fields}
\author{\"{O}. A\c{c}{\i}k $^{1}$}
\email{ozacik@science.ankara.edu.tr}
\author{\"{U}. Ertem $^{1}$}
\email{uertem@science.ankara.edu.tr}
\author{M. \"{O}nder $^{2}$}
 \email{onder@hacettepe.edu.tr}
 \author{A. Ver\c{c}in $^{1}$}
 \email{vercin@science.ankara.edu.tr}
\address{$^{1}$ Department of Physics, Ankara University, Faculty of Sciences,
06100, Tando\u gan-Ankara, Turkey\\
$^{2}$ Department of Physics Engineering, Hacettepe University,
06800, Beytepe-Ankara, Turkey.}

\date{\today}

\begin{abstract}

Killing-Yano one forms (duals of Killing vector fields) of a class
of spherically symmetric space-times characterized by four functions
are derived in a unified and exhaustive way. For well-known
space-times such as those of Minkowski, Schwarzschild,
Reissner-Nordstr{\o}m, Robertson-Walker and several forms of de
Sitter, these forms arise as special cases in a natural way. Besides
its two well-known forms, four more forms of de Sitter space-time
are also established with ten independent Killing vector fields for
which four different time evolution regimes can explicitly be
specified by the symmetry requirement. A family of space-times in
which metric characterizing functions are of the general form and
admitting  six or seven independent Killing vector fields is
presented.
\end{abstract}


\maketitle

\section{Introduction}

Defining relations of Killing-Yano (KY) and conformal KY-forms  are
natural generalizations of Killing 1-forms and conformal Killing
1-forms. The latter are the dual of Killing and conformal Killing
vector fields whose flows generate, respectively, local isometries
and local conformal isometries of the metric in (pseudo)Riemannian
geometry \cite{Benn-Tucker,Thirring}. Although they are not related
to the isometries of the metric these higher rank generalizations
have attracted increasing interest in various fields of physics and
modern mathematics as well as in some related fields. Generally
speaking, many interesting properties of a space-time are intimately
connected with the existence of (conformal) KY-forms admitted by the
corresponding metric. More specifically; the determination of
KY-forms of a given metric, classification of space-times admitting
KY-forms, analysis of the algebraic structures of these forms as
well as specification of the symmetry algebra and related conserved
quantities of the Dirac and related equations in a given curved
background have gained increasing significance. Equally important
objects (not considered in this study) are the totally symmetric
Killing tensors and their conformal generalizations (see
\cite{Dietz1, Dietz2,Benn-son} and references therein).

KY-forms play a prominent role in a unified description of null and
non-null shear-free congruences \cite{Benn} and in the search of
force-free fields (divergence-free eigenvectors of the curl operator
with position dependent eigenvalues) mostly encountered in
astrophysics and fluid dynamics literatures
\cite{Benn-Kress0,Benn-Kress3,Kress}. A tangent vector which
generates conformal transformations on its orthogonal complement is
said to be the generator of shear-free congruence for its integral
curve. Shear-free equation is a generalization of defining equation
of conformal Killing equation. Together with Clifford calculus,
KY-forms also provide efficient means in analyzing elliptic
operators and the Dirac operator and in the further classification
of (pseudo)Riemannian manifolds \cite{Benn-Tucker,Semmelman1}. While
conformal KY-forms take part in symmetry operators for the massless
Dirac equation \cite{Benn-Charlton,Benn-Kress3}, KY-forms are
indispensable in constructing first order symmetries of the massless
as well as massive Dirac equation in a curved space-time
\cite{Benn-Kress1}. KY-forms are also necessary for the symmetries
of the K\"{a}hler equation \cite{Benn-Tucker,Benn-Kress2}.

Longstanding interest in KY-forms largely stems from their constant
use in general relativity and especially from their role in
constructing conserved quantities in a number of ways. The studies
of Penrose and his collaborators \cite{Walker,Penrose}, who have
shown how the existence of a KY 2-form explains Carter's result on
the integrability of the geodesic equation in Kerr background,
constitute a stepping-stone in this context \cite{Carter1,Carter2}.
The fact that any KY p-form provides a quadratic first integral of
the geodesic equation is by now a well-established particular result
of the fact that the interior derivative of any KY p-form with
respect to the tangent vector field of any geodesic remains parallel
along the geodesic. More generally, any KY (p+1)-form is associated
with a symmetric bilinear form, which is nothing more than the
Killing tensor generalizing the so-called Stackel-Killing tensor
that corresponds to a KY 2-form as first recognized by Penrose and
Floyd. For more in this context, refer to \cite{Semmelman1}.

On the other hand, as every KY form is co-closed, the Hodge dual of
any KY $p$-form is directly associated with a conserved quantity. In
the case of KY 1-forms, two types of conserved currents can be
defined. For Ricci-flat space-times the Hodge dual of the exterior
derivative of a KY 1-form $\omega$ is conserved, where
$^{\ast}d\omega$ is known as the Komar form \cite{Benn-Tucker}.
Secondly, the current $j=i_{X^{a}}\omega\wedge^{\ast^{-1}} G_a$
defined in terms of the Einstein $3$-forms $G_a$, and KY 1-form
$\omega$ is also conserved. Here, $^{\ast}$ and $^{\ast^{-1}}$
represent the Hodge map and its inverse, $\wedge$ denotes exterior
multiplication, $d$ and $i_{X}$ stand for exterior and interior
derivatives (with respect to vector field $X$).  It has recently
been shown how the KY-forms of the flat space-time can be used to
construct new, conserved gravitational charges for transverse
asymptotically flat \cite{Kastor} as well as for asymptotically anti
de Sitter space-times \cite{Cebeci}. These studies present another
way of constructing a conserved current by taking a particular
linear combination of wedge products of interior derivatives of the
KY p-form and curvature characteristics of the underlying manifold.
For KY 1-forms, this reduces to the usual current obtained from the
Einstein $3$-form.

The main purpose of this and the accompanying paper \cite{ozumav1}
is to develop, by directly starting from the KY-equation,
\begin{eqnarray}
\nabla_{X_{a}}\omega_{(p)} =\frac{1}{p+1}
i_{X_{a}}d\omega_{(p)}\;,\quad p=1,2,3
\end{eqnarray}
a constructive method which makes it possible to generate all KY
forms for a large class of spherically symmetric space-times in a
unified and exhaustive way. Here $\nabla_{X}$ stands for the
covariant derivative with respect to the vector field $X$. It should
be noted that the KY 0-form $\omega_{(0)}$ can be any function,
$\omega_{(1)}$ is the dual of a Killing vector field and
$\omega_{(n)}$ is a constant (parallel), that is, it is a constant
multiple $\omega_{(n)}=az$ of the volume form $z$ (for the
orthonormal frame given below $z=e^{0123}$). The goal of our study
is achieved by solving a coupled set of first order partial linear
differential equations for the component functions of the KY forms
$\omega_{(p)}$ in each case. The number of independent equations
that have to be solved for $ p=1,2,3$ are $10,\;18$ and $16$,
respectively. To obtain the most general solutions in an exhaustive
way, there are also $24, 36$ and $24$ integrability conditions that
must be carefully examined at the outset. We first solve a suitable
set of the integrability conditions by which the set of solutions
naturally branches into cases and subcases.

We shall mainly use the notation of \cite{Benn-Tucker} and adopt the
following conventions and terminology. The underlying base manifold
is supposed to be a $4$-dimensional ($4D$) pseudo-Riemannian
manifold with the metric tensor $g$ having Lorentzian signature
$(-+++)$ such that $g=-e^{0}\otimes e^{0}+e^{1}\otimes
e^{1}+e^{2}\otimes e^{2}+e^{3}\otimes e^{3}$ in a local orthonormal
co-frame $\{e^{a}\}$. By choosing the co-frame basis
\begin{eqnarray}
 e^0 &=& H_{0}dt\;,\qquad
 e^1=T H_{1}dr\;,\nonumber\\
 e^2 &=& T H_{2}d\theta\;,\quad
 e^3=T H_{2}\sin\theta d\varphi \;,\nonumber
\end{eqnarray}
a class of spherically symmetric metrics that will be considered can
be parameterized by
\begin{eqnarray}
T=\exp(\lambda(t))\;,\quad H_{j}=H_{j}(r)\;;\quad j=0,1,2\;\nonumber
\end{eqnarray}
which henceforth will be referred to as the (metric) coefficient
functions. Here $(t,r,\theta,\varphi)$ specifies a local polar
space-time chart with the usual range of variations. Whenever
necessary, the range of $r$ can be bounded to keep the coefficient
functions real. A generic property of this kind of metric is
invariance under the transformation of the spatial rotation group
$SO(3)$. If $g$ admits a time-like Killing vector field $K_0$, it is
termed stationary and if, in addition, $K_0$ is orthogonal to a
family of space-like hypersurfaces it is termed static. The dual
tangent frame basis $\{X_{a}\}$ of the co-frame basis $\{e^{a}\}$
are,
\begin{eqnarray}
X_{0} &=& \frac{1}{H_{0}}\partial_{t}\;,\qquad
X_1=\frac{1}{TH_{1}}\partial_{r}\;,\nonumber\\
X_2 &=& \frac{1}{T H_{2}}\partial_{\theta}\;,\quad X_3=\frac{1}{T
H_{2}\sin\theta}\partial_{\varphi}\;,\nonumber
 \end{eqnarray}
where $e^{a}(X_b)=\delta^{a}_{b}$ and $\partial_x=\partial/\partial
x$. The metric dual of a vector field $X$ will be denoted by
$\tilde{X}$ such that $\tilde{X}(Y)=g(X,Y)$ for any vector field
$Y$. The torsion-free connection $1$-forms $\omega_{ab}$ for this
class of metrics are well-known, and can be presented in the
following antisymmetric matrix-valued $1-$form:
\begin{eqnarray}
(\omega_{ab})=\frac{1}{T}\left(
\begin{array}{cccc}
0 &  &  & \\
\frac{h_{0}}{H_{1}}e^{0}+\frac{\dot{T}}{H_{0}}e^{1} & 0 &  & \\
\frac{\dot{T}}{H_{0}}e^{2} & \frac{h_{2}}{H_{1}}e^{2} & 0 & \\
\frac{\dot{T}}{H_{0}}e^{3} & \frac{h_{2}}{H_{1}}e^{3} &
\frac{\cot\theta}{H_{2}}e^{3} & 0
\end{array}
\right)\;,
\end{eqnarray}
where we have used the abbreviations
$\dot{T}=dT/dt,\;dH(r)/dr=H^{\prime}$ and
$h_{j}=H^{\prime}_{j}/H_{j}$. We shall always use prime and over-dot
to denote, respectively, the $r$-derivation and $t$-derivation of a
function which depends only on $r$ and $t$. The partial derivative
of a map $U$ of several variables with respect to $x$ will be
denoted by $U_x$, and of a rational map or function $U/V$ by
$\partial_x(U/V)$. The following matrix-valued $1$-form is helpful
in carrying out the calculations
\begin{eqnarray}
(\nabla_{X_{a}}e^{b})=-\frac{1}{T}\left(
\begin{array}{cccc}
\frac{h_{0}}{H_{1}}e^{1} \;& \frac{h_{0}}{H_{1}}e^{0} \;& 0 & 0 \\
\frac{\dot{T}}{H_{0}}e^{1} \;& \frac{\dot{T}}{H_{0}}e^{0} \;& 0 & 0 \\
\frac{\dot{T}}{H_{0}}e^{2} \;& -\frac{h_{2}}{H_{1}}e^{2}\; &
\frac{\dot{T}}{H_{0}}e^{0}+\frac{h_{2}}{H_{1}}e^{1} & 0 \\
\frac{\dot{T}}{H_{0}}e^{3} \;& -\frac{h_{2}}{H_{1}}e^{3} \;&
-\frac{\cot\theta}{H_{2}}e^{3} & \frac{\dot{T}}{H_{0}}e^{0}
+\frac{h_{2}}{H_{1}}e^{1} +\frac{\cot\theta}{H_{2}}e^{2}
\end{array}
\right)\;.
\end{eqnarray}
The corresponding curvature $2$-forms are presented in Appendix B.

All the well-known spherically symmetric space-times such as the
Minkowski, Schwarzschild, Reissner-Nordstr{\o}m, Robertson-Walker
and the six various forms of de Sitter models fall within the class
of the considered metric as special cases and this provides the
opportunity to give complete lists of their KY forms. As a
particular result, we have found a completely solvable nonlinear
ordinary differential equation $T^{2}\partial_t(\dot{T}/T)=\ell$,
where $\ell$ is constant, characterizing five different
time-dependent types of de Sitter space-time in a unified way. This
fact also enables us to give explicit expressions of their KY-forms
in a unified and exhaustive way. As the first part of our study, the
present paper is entirely devoted to the unified generation of all
Killing vector fields for the considered class of space-times. KY
two and three forms are taken up in the next paper \cite{ozumav1}.

Although KY-forms have been the subject of relatively recent active
research, Killing vectors have been so intensely investigated that
the following original points which hold for the KY-forms as well
are worth emphasizing. (i) As has been mentioned above, there exist
five well-known space-times that are covered by the considered class
of metrics such that one of them (de Sitter) consists of six
different types. Killing vector fields of these metrics  are usually
handled as case-by-case studies in scattered references. Our unified
generation may remedy many inconveniences such as notational
incompatibilities, proper range problems related to coordinates and
relationships between cases. Moreover, one can explicitly observe
the emergence of each case from the variations of the metric
characterizing coefficients. (ii) Derivations of Killing vector
fields for some types of de Sitter space-time from a five
dimensional embedding flat manifold is geometrically very appealing
and more inspiring physically (for a review of de Sitter spaces see
\cite{Hawking,Moschella} and for recent interest see \cite{Randono}
and references therein). But the exact number of possible types
naturally emerges from our study, and we identify an exactly
solvable equation that determines this number by the number of its
possible solutions. (iii) Our approach is exhaustive in the sense
that all of the possible Killing vector fields of a given space-time
can be completely determined from our approach so long as its metric
belongs to the considered class. This fact makes it possible to
reach decisive, or at least conclusive statements about a particular
problem in which KY-forms are involved. We do not go into the detail
of all possible cases but point out sufficiently symmetric cases,
and have given some details of a particular case having six
independent Killing vector fields that, as far as we know, does not
appear in the literature. This example is also  worth mentioning in
light of the fact that its symmetry algebra changes drastically when
a seemingly unimportant integration constant is changed.

\section{KY 1-Forms: Defining Equations}

A $1$-form is a KY $1$-form if and only if it is the metric dual of
a Killing vector field. This is equivalent to the fact that it
satisfies equation (1) for $p=1$. For the components of
\begin{eqnarray}
\omega_{(1)}=\alpha e^0+\beta e^1+\gamma e^2+\delta e^3\;,\nonumber
 \end{eqnarray}
the KY equation gives, in view of (2) and (3), sixteen equations of
which the following ten
\begin{eqnarray}
\alpha_{t}&=&\frac{H^{\prime}_{0}}{TH_1}\beta\;,\;\quad\quad\quad
\beta_{r} =\dot{T}\frac{H_{1}}{H_0}\alpha\;,\nonumber\\
T^{2}\partial_{t}\frac{\beta}{T}&=&-
\frac{H^{2}_{0}}{H_1}\partial_{r}\frac{\alpha}{H_0}\;,\;\;\quad
\beta_{\theta}=-\frac{H^{2}_2}{H_1}\partial_{r}\frac{\gamma}{H_{2}}\;,\nonumber\\
T^{2}\partial_{t}\frac{\gamma}{T}&=&-\frac{H_{0}}{H_2}\alpha_{\theta}
\;,\;\;\;\quad\quad
\beta_{\varphi}=-\frac{H^{2}_2}{H_1}\sin\theta\partial_{r}
 \frac{\delta}{H_{2}}\;,\\
T^{2}\partial_{t}\frac{\delta}{T}&=&
-\frac{H_{0}}{H_2\sin\theta}\alpha_{\varphi}\;,\;\;\;
\gamma_{\theta} =\dot{T}\frac{H_2}{H_0}\alpha-
 \frac{H'_{2}}{H_1}\beta\;,\nonumber\\
\delta_{\varphi}&=&
 \sin^{2}\theta\partial_{\theta}\frac{\gamma}{\sin\theta}
 \;,\quad \gamma_{\varphi} =-\sin^{2}\theta\partial_{\theta}\frac{\delta}{\sin\theta}
\nonumber
\end{eqnarray}
are independent. Although it appears difficult to solve this coupled
set of first order partial differential equations directly, an
exhaustive treatment of the problem with many classes of solutions
is possible.

At first to gain an initial insight into the above set of equations,
let us look at some obvious solutions. We can immediately see that
when all coefficient functions but $\alpha$ are zero, $T$ must be
constant and $\alpha=H_0$. When $\beta$ alone is nonzero and
proportional to $T$, the conditions
$H^{\prime}_{0}=0=H^{\prime}_{2}$ must be fulfilled. The other two
cases, in which only $\gamma$ or only $\delta$ is different from
zero and proportional to $TH_2$, are forbidden by the last two
equations of (4). But it is easy to verify that
\begin{eqnarray}
\alpha=0=\beta\;,\quad\gamma=TH_{2}g\;,\quad
\delta=TH_{2}(a_{3}\sin\theta+\cos\theta g_{\varphi})\;,\nonumber
\end{eqnarray}
is a solution, such that $g(\varphi)$ satisfies
$g_{\varphi\varphi}+g=0$ and there is no additional constraint. We
thus have, for $g=a_1\sin\varphi-a_2\cos\varphi$, the following
three linearly independent KY 1-forms :
\begin{eqnarray}
\tilde{K}_{1}&=& TH_2(\sin\varphi e^{2}+\cos\theta\cos\varphi e^{3})
 \;,\nonumber\\
\tilde{K}_{2}&=&TH_2( -\cos\varphi e^{2}+\cos\theta\sin\varphi
e^{3})
 \;,\\
 \tilde{K}_{3}&=&-TH_2\sin\theta e^{3}\;.\nonumber
\end{eqnarray}
The corresponding rotational Killing vector fields
\begin{eqnarray}
K_{1}&=&
\sin\varphi\partial_{\theta}+\cot\theta\cos\varphi\partial_\varphi
 \;,\nonumber\\
K_{2}&=&
-\cos\varphi\partial_{\theta}+\cot\theta\sin\varphi\partial_\varphi
 \;,\\
 K_{3}&=&
-\partial_{\varphi}\;,\nonumber
\end{eqnarray}
are the well-known generators of the $so(3)$-algebra:
$[K_{i},K_{j}]=\varepsilon_{ij}^{\;\;\;k}K_{k}$. As there is no
constraint on these solutions, they must appear in every case
independent of the specific form of the functions characterizing the
metric. This is a typical characteristic of spherical symmetry.

When $\dot{T}=0,\;\omega_0=H_0 e^{0}$ is a KY $1$-form which
corresponds to the time-like Killing vector field
$K_0=\tilde{\omega}_0=-\partial_t$, and it can be combined with the
above $so(3)$ solutions. In fact for $\alpha=H_0,\;\beta=0$ and
$\dot{T}=0$, the above $4D$ algebra is the dual of the most general
solution of equations (4). Indeed in such a case, the first eight
equations of (4) imply that $\gamma=TH_{2}G(\varphi)$ and
$\delta=TH_{2}D(\theta,\varphi)$ and then the last two equations of
(4) yield
\begin{eqnarray}
D_\varphi=-\cos\theta G\;,\quad
G_\varphi=-\sin\theta D_\theta+\cos\theta D\;.\nonumber
\end{eqnarray}
From the derivation of the second equation with respect to
$\varphi$, we obtain $G_{\varphi\varphi}+G=0$ in view of the first
equation. On the other hand, integration of the first equation gives
$D=\cos\theta G_\varphi+f(\theta)$ which, upon substituting it into
the second, gives $f_\theta=\cot\theta f$ whose integral is
$f=a_3\sin\theta$. In the case of $H_2=r$, such space-times with
$4D$ symmetry algebra include two physically important examples: the
Reissner-Nordstr\"{o}m (RN) and its special case Schwarzschild
space-times, for which the other coefficient functions are given in
Table I. It should be emphasized that the explicit forms of $H_0$
and $H_1$ are derived from the physical requirements, namely from
the Einstein equations in the Schwarzschild case.

For a general consideration, it turns out to be convenient to look
for the solutions in the set of the solutions of some integrability
conditions. This will also provide us with the necessary  means to
generate other sets of solutions.

\section{Integrability Conditions and Their Solutions}

For $x=\alpha,\beta$ we have the integrability conditions
\begin{eqnarray}
\partial_{\varphi}\partial_\theta(\frac{x}{\sin\theta})=0\;,\quad
x_{\varphi\varphi}=\sin^{3}\theta\partial_{\theta}
\frac{x_{\theta}}{\sin\theta}\;,
\end{eqnarray}
that follow from $\delta_{r\theta}=\delta_{\theta
r},\delta_{rt}=\delta_{tr},\delta_{t\theta}=\delta_{\theta t},
\delta_{t\varphi}=\delta_{\varphi t}$ and
$\gamma_{r\varphi}=\gamma_{\varphi r},
\gamma_{\theta\varphi}=\gamma_{\varphi\theta}$. There are two
additional conditions for $\alpha$ and four conditions for $\gamma$
and $\delta$:
\begin{eqnarray}
\partial_{r}\partial_\theta(\frac{\alpha}{H_{2}})&=&0=
\partial_{r}\partial_\varphi(\frac{\alpha}{H_{2}})\;,\nonumber\\
\partial_{r}\partial_\theta(\frac{\delta}{H_{2}})&=&0=
\partial_{t}\partial_\theta(\frac{\delta}{T})\;,\\
\partial_{t}\partial_r(\frac{\delta}{TH_{0}})
&=&0=
\partial_{t}\partial_r(\frac{\gamma}{TH_{0}})\;.\nonumber
\end{eqnarray}
The first two conditions of (8) can be checked from
$\alpha_{r\theta}=\alpha_{\theta r}$ and
$\alpha_{r\varphi}=\alpha_{\varphi r}$. The second row of (8)
follows from $\alpha_{\theta\varphi}=\alpha_{\varphi\theta},
\beta_{\theta\varphi}=\beta_{\varphi\theta}$ and the last row can be
seen from $\beta_{t\varphi}=\beta_{\varphi t}$ and
$\beta_{r\theta}=\beta_{\theta r}$. The following two equations can
be easily verified from the last row of (4)
\begin{eqnarray}
y_{\varphi\varphi}+y=-\sin^{3}\theta\partial_{\theta}
\frac{y_{\theta}}{\sin\theta}\;,
\end{eqnarray}
for $y=\gamma,\delta$. There are $24$ integrability conditions but
only $20$ of them are independent, and the twelve shown above are
sufficient for a unified and exhaustive investigation.

\subsubsection{The General Forms of $\alpha$ and $\beta$}

In terms of the functions $f=f(t,r,\varphi), g=g(t,r,\theta)$ the
first equation of (7) implies
\begin{eqnarray}
x_\varphi=\sin\theta f\;,\quad x_\theta=\cot\theta x+g\;,\nonumber
\end{eqnarray}
and from the second equation of (7) we obtain $x=\sin^{2}\theta
g_\theta-\sin\theta f_\varphi$. On substituting this solution into
the above $x_\varphi$ and $x_\theta$ equations we arrive at
\begin{eqnarray}
f_{\varphi\varphi}+f=0\;,\;g=\sin\theta
\partial_\theta(\sin\theta g_{\theta})\;,\nonumber
\end{eqnarray}
whose general solutions can be written, with $f_i=f_i(t,r)$ and
$g_{i}=g_{i}(t,r)$, as
\begin{eqnarray}
f=-f_{1}\sin\varphi-f_{2}\cos\varphi\;,\quad g=-g_1\cot\theta
-g_2\frac{1}{\sin\theta}\;.\nonumber
\end{eqnarray}
Here the minus signs are used for convenience. The general solution
for $x$ is
\begin{eqnarray}
x=g_{1}+g_2\cos\theta+\sin\theta(f_1\cos\varphi-f_2\sin\varphi)
\;.\nonumber
\end{eqnarray}
In terms of the functions $A_i=A_i(t,r)$ and $B_i=B_i(t,r)$ let us
define
\begin{eqnarray}
\sigma^{A}&=&A_1\cos\varphi-A_2\sin\varphi\;,\quad
\sigma^{B}=B_1\cos\varphi-B_2\sin\varphi\;, \nonumber\\
A&=&\sin\theta\sigma^{A}+A_{3}\cos\theta\;,\quad\quad
B=\sin\theta\sigma^{B}+B_{3}\cos\theta\;.\nonumber
\end{eqnarray}
Since the functions $f_i$ and $g_i$ are, in general, different for
$\alpha$ and $\beta$, we can write their general forms very
concisely as
\begin{eqnarray}
\alpha=U+A\;,\quad \beta=V+B\;,
\end{eqnarray}
where $U$ and $V$ depend, like the $g_1$ term of $x$, on $t$ and
$r$. Note that $A$ and $B$ depend on all of the coordinates and they
satisfy the relations
\begin{eqnarray}
A_{\theta\theta}+A=0=B_{\theta\theta}+B\;,\quad
\sigma^{A}_{\varphi\varphi}+\sigma^{A}=0=\sigma^{B}_{\varphi\varphi}+\sigma^{B}\;.
\end{eqnarray}

In the case of $x=\alpha$, the first two conditions of (8) imply
that the functions characterizing $A$ are proportional to $H_{2}$
and these enable us to write $A=H_2\xi$ such that
\begin{eqnarray}
\xi=u\cos\theta+\sin\theta(v_1\cos\varphi-v_2\sin\varphi)\;,
\end{eqnarray}
where $u,v_i$ depend only on $t$.

\subsubsection{The General Forms of $\gamma$ and $\delta$}

When $\alpha$ and $\beta$ given by (10) are substituted into the
$\gamma_\theta$-equation of (4), we obtain
\begin{eqnarray}
\gamma_\theta=(\dot{T}\frac{H_2}{H_0}U-\frac{H_{2}^{\prime}}{H_{1}}V)+
\dot{T}\frac{H_2}{H_0}A-\frac{H_{2}^{\prime}}{H_{1}}B\;.\nonumber
\end{eqnarray}
For well-defined $\gamma$ solutions, the first term at the right
hand side must be zero:
\begin{eqnarray}
\dot{T}U=H_{0}PV\;,\quad (P=\frac{H_2^{\prime}}{H_1H_2})\;.
\end{eqnarray}
(Since the mentioned term is independent of $\theta$ it would lead,
upon integration, to a $\gamma$ solution which linearly depends on
$\theta$. In fact this condition results when the above
$\gamma_\theta$ is used in eq. (9) of $\gamma$.) Then, in view of
eq.(11), the $\gamma_\theta$-equation can be integrated to
\begin{eqnarray}
\gamma=-(\dot{T}\frac{H_2}{H_0}A_{\theta}-\frac{H_{2}^{\prime}}{H_{1}}B_{\theta})
+TH_2g\;,\;\;g_{\varphi\varphi}+g=0\;,
\end{eqnarray}
where the $\theta$-independent term $G=TH_2g(\varphi)$ comes from
integration, and its form can be easily verified by substituting
$\gamma$ in the $\alpha_\theta$ and $\beta_\theta$-equations of (4).
Indeed, in the first case we get
\begin{eqnarray}
T^{2}\partial_t(\frac{\dot{T}}{T}A_{\theta})-\frac{H_{0}^{2}}{H_{2}^{2}}A_{\theta}
=T^{2}H_0P\partial_t\frac{B_{\theta}}{T}\;,
\end{eqnarray}
in addition to $\partial_t(G/T)=0$. In the second case, we obtain
\begin{eqnarray}
\dot{T}\partial_r(\frac{A_{\theta}}{H_0})=
\frac{H_{1}}{H_{2}^{2}}B_{\theta}+\partial_r(PB_{\theta})\;,
\end{eqnarray}
and $\partial_r(G/H_2)=0$. These imply that $G$ is of the form
$G=TH_2g(\varphi)$ and by the integrability condition (9) for
$\gamma$, we see that $g$ must satisfy the equation given by (14).

Having determined the general form of $\gamma$, we now use it in the
last two equations of (4) to determine the form of $\delta$. The
$\delta_\varphi$-equation of (4) directly gives
\begin{eqnarray}
\delta_\varphi=\dot{T}\frac{H_2}{H_0}\sigma^{A}-\frac{H_{2}^{\prime}}{H_{1}}\sigma^{B}
-TH_2\cos\theta g\;,\nonumber
\end{eqnarray}
which can be integrated to
\begin{eqnarray}
\delta=-\dot{T}\frac{H_2}{H_0}\sigma^{A}_\varphi+\frac{H_{2}^{\prime}}{H_{1}}
\sigma^{B}_\varphi+TH_2\cos\theta g_\varphi+aTH_2\sin\theta\;,
\end{eqnarray}
where $a$ is an integration constant and the $\varphi$-independent
term $D=aTH_2\sin\theta$ again comes from integration, whose
explicit form is determined from $\alpha_\varphi,\;\beta_\varphi$
and the last two equations of (4). Indeed, the
$\alpha_\varphi$-equation of (4) states that $D/T$ must be
independent of $t$ and that the following condition must be
satisfied :
\begin{eqnarray}
T^{2}\partial_t(\frac{\dot{T}}{T}\sigma^{A}_{\varphi})-
\frac{H_{0}^{2}}{H_{2}^{2}}\sigma^{A}_{\varphi}
=T^{2}H_0P\partial_t\frac{\sigma^{B}_{\varphi}}{T}\;.
\end{eqnarray}
On the other hand, the $\beta_\varphi$-equation of (4) states that
$D/H_2$ must be independent of $r$, and that the following condition
must be satisfied :
\begin{eqnarray}
\dot{T}\partial_r\frac{\sigma^{A}_{\varphi}}{H_0}=\frac{H_1}{H_2^{2}}
\sigma^{B}_{\varphi}+\partial_r(P\sigma^{B}_{\varphi})\;.
\end{eqnarray}
As a particular result, we have $D=TH_2f(\theta)$ and when the found
forms of $\gamma$ and $\delta$ are used in the last equation of (4),
we obtain $f=a\sin\theta$. Note that the conditions (18) and (19)
are contained in (15) and (16).

As an intermediate result, all the metric coefficient functions have
been specified in terms of seven functions $U,A_i$ and $B_i$, which
can be determined from the first three equations of (4) and the
conditions found in (13), (15) and (16). Therefore, seven equations
of (4) have been analyzed, and the first three equations and
conditions (15) and (16) remain to be solved. Note that the last
terms of $\gamma$ and $\delta$ correspond to the $so(3)$ solutions.
The rest of the investigation is entirely devoted to the
specification of additional symmetries.

\subsubsection{The Clustering of Solutions}

We shall now substitute the solutions (10) and (12) into the first
three equations of (4) to specify the unknown functions. The first
three equations of (4) give the following three equations for $U$
\begin{eqnarray}
U_t&=&\frac{\dot{T}}{T}\frac{H_0^{\prime}H_2}{H_0H^{\prime}_2}U\;,\nonumber\\
\partial_r(\frac{U}{H_0P})&=&H_1\frac{U}{H_0}\;,\\
T^{2}\partial_t(\frac{\dot{T}}{T}U)&=&-\frac{H_0^{3}P}{H_1}\partial_r\frac{U}{H_0}\;,\nonumber
\end{eqnarray}
and the following three equations for $B$
\begin{eqnarray}
B=TL\xi_{t}\;,\quad B_r=\dot{T}\frac{H_1H_2}{H_0}\xi\;,\quad
T^{2}\partial_t\frac{B}{T}=-\frac{H_0^{2}}{H_1}(\frac{H_2}{H_0})^{\prime}\xi\;,
\end{eqnarray}
where we have utilized relation (13) and the function $L=L(r)$ is
defined by
\begin{eqnarray}
L=\frac{H_1H_2}{H_{0}^{\prime}}\;.
\end{eqnarray}

As is obvious from the equations in this subsection, from here on
the analysis critically depends on the derivatives of $T, H_0$ and
$H_2$. Since $H_2^{\prime}$ is different from zero in all physically
important space-times, we shall assume this to be the case
throughout the paper. This means that the function $P$ is different
from zero. The cases $H_0^{\prime}=0$ and $\dot{T}=0$ will be
considered as particular cases in the last three sections. In the
next section, we proceed to look for solutions for which both
$H_0^{\prime}$ and $\dot{T}$ can be different from zero.

\section{Three classes of solutions}

Using the first equation of (21) in the other two equations of (21),
we obtain two equations which accept separation of variables such
that
\begin{eqnarray}
m\xi_t=\frac{\dot{T}}{T}\xi\;,\quad T^{2}\xi_{tt}=-m_0\xi\;,
\end{eqnarray}
provided that $m$ and $m_0$ defined by
\begin{eqnarray}
m=L^{\prime}\frac{H_0}{H_1H_2}\;,\quad
m_0=\frac{H^{2}_0}{LH_1}(\frac{H_2}{H_{0}})^{\prime}\;,
\end{eqnarray}
are constants. When one side of a separable equation depends
entirely on the metric coefficient functions, we shall use the
letters $k,l$ and $m$ for separation constants. Other integration
constants will be denoted by the letters $a,b$ and $c$.

In view of the ansatz $U=H_0Y(r)f(t)$ the first two equations of
(20) transform to
\begin{eqnarray}
f_t=k\frac{\dot{T}}{T}f\;,\quad (\frac{Y}{P})^{\prime}=H_1Y\;,
\end{eqnarray}
where the metric constant $k$ is defined by
\begin{eqnarray}
k=\frac{H^{\prime}_0H_2}{H_0H_2^{\prime}}\;.
\end{eqnarray}
On the other hand from the last equation of (20), with the same
ansatz, we get
\begin{eqnarray}
T^{2}\partial_t(\frac{\dot{T}}{T}f)=k_0f\;,\quad
\frac{Y^{\prime}}{Y}=-k_0\frac{H_1}{H_0^{2}P}\;,
\end{eqnarray}
where $k_0$ is a separation constant. Two equations of (25) can be
easily integrated to
\begin{eqnarray}
f=c_1T^{k}\;,\quad Y=c_2H_2P\;.
\end{eqnarray}
Thus $U=cH_0H_2PT^{k}$ where $c_i$ are integration constants and
$c=c_1c_2$. When these solutions are substituted into (27) we see
that $T$ must satisfy
\begin{eqnarray}
T\ddot{T}+(k-1)\dot{T}^{2}=k_0\;,
\end{eqnarray}
and $k_0$  must be the metric constant
\begin{eqnarray}
k_0=-\frac{H_0^{2}}{H_1H_2}(\frac{H_2^{\prime}}{H_1})^{\prime}\;.
\end{eqnarray}

Noting that equations (20) and (21) are linear in $U$ and $B$,
depending on the values of $m$ one can distinguish three classes of
solutions for which both $H_0^{\prime}$ and $\dot{T}$ can be
different from zero. These can be characterized as follows
\begin{eqnarray}
(A)\;\; m=0\;,\quad (B)\;\;m\neq 0\;;\;\;U=0=V\;,\quad (C)\;\;m\neq
0\;.\nonumber
\end{eqnarray}
The cases (i) $\dot{T}=0,\;H_0^{\prime}\neq 0$, (ii)
$\dot{T}\neq0,\;H_0^{\prime}=0$ and (iii) $\dot{T}=0=H_0^{\prime}$
will be considered separately in the last three sections.

\subsection{ $m=0$ Solutions}

When $m$ is zero $L$ is a nonzero constant and we have
${H_{0}^{\prime}}L=H_1H_2$. In that case, equations given by (23)
imply that $\xi=0=B$ and that $m_0$ need not be a constant. The
conditions (15) and (16) are both trivially satisfied for this case
and hence, $\gamma$ and $\delta$ solutions correspond only to the
$so(3)$ solutions. Only the metric constants $k$ and $k_0$ are
defined in that case and, provided that $T$ satisfies equation (29),
we have the following solutions for $\alpha$ and $\beta$:
\begin{eqnarray}
\alpha=U=cT^{k}H_0\frac{H_2^{\prime}}{H_1}\;,\quad
\beta=V=c\dot{T}T^{k}H_2 \;,
\end{eqnarray}
which determine the Killing vector field
\begin{eqnarray}
X=-\frac{H_2^{\prime}}{H_1}T^{k}\partial_t+\dot{T}T^{k-1}\frac{H_2}{H_1}\partial_r
\;,
\end{eqnarray}
commuting with the $so(3)$ vector fields.

\subsection{$m\neq 0\;{\rm and}\;U=0=V$ Solutions}

When $m$ is different from zero, the first equation of (23) can be
easily solved to be
\begin{eqnarray}
\xi=T^{1/m}\zeta\;,\quad
\zeta=a_1\cos\theta+\sin\theta(a_2\cos\varphi-a_3\sin\varphi)\;,
\end{eqnarray}
where $a_i$ are integration constants. The second equation of (23)
then implies that
\begin{eqnarray}
T^{(2m-1)/m}\partial_t(\dot{T}T^{(1-m)/m})=-mm_0\;.
\end{eqnarray}
This equation can equivalently be read as
\begin{eqnarray}
(1-m)\dot{T}^{2}+mT\ddot{T}=-m^{2}m_0\;,
\end{eqnarray}
or more concisely, in terms of $K=T^{1/m}$, as
\begin{eqnarray}
\ddot{K}=-m_0 K^{1-2m} \;.
\end{eqnarray}
The corresponding $\alpha,\beta,\gamma$ and $\delta$ solutions are
as follows:
\begin{eqnarray}
\alpha &=& T^{1/m}H_2\zeta\;,\quad \beta=WL\zeta\;,\nonumber\\
\gamma&=&\frac{1}{m_{0}}WH_0\zeta_\theta+TH_2 g\;, \quad
(g_{\varphi\varphi}+g=0) \\
\delta &=&\frac{1}{m_{0}}WH_0\sigma_\varphi+TH_2(a\sin\theta+
\cos\theta g_\varphi)\;,\nonumber
\end{eqnarray}
where $W=\dot{T}T^{1/m}/m=\dot{K}T$ and
\begin{eqnarray}
\sigma^{A}=T^{1/m}H_2\sigma\;,\quad \sigma^{B}=WL\sigma\;,\quad
\sigma=a_2\cos\varphi-a_3\sin\varphi\;.\nonumber
\end{eqnarray}
In view of (24) and equation (35); it is easy to verify that while
condition (15) yields,
\begin{eqnarray}
m_{0}(\frac{H_2^{\prime}}{H_{0}^{\prime}}-m\frac{H_2}{H_{0}})=\frac{H_0}{H_{2}}
\;,
\end{eqnarray}
for $A=T^{1/m}H_2\zeta$ and $B=WL\zeta$, condition (16) yields an
equation that is just the derivative of (38). In writing $\gamma$
and $\delta$ of (37) the condition (38) has been used.

\subsection{ $m\neq 0$ Solutions}

In this case, solutions are just a combination of the above two
classes: $\gamma$ and $\delta$ solutions are as in equations (37)
but to the $\alpha$ and $\beta$ solutions of (37), one must add that
given by (31). Equation (38) also hold in this case. However, there
are four metric constants $m,m_0,k$ and $k_0$, and $T$ must satisfy
both the equations (29) and (34) (or equivalently (35) or (36)). A
detailed investigation of the four metric constants presented in
Appendix A shows that for $H^{\prime}_2$ to be nonzero, the
following conditions must be satisfied:
\begin{eqnarray}
km=1\;,\quad m=1+m_0\ell^{2}\;,\nonumber
\end{eqnarray}
where $\ell$ is another metric constant. In fact, from equations
(29), (35) and the above relation, we also get $k_0=-mm_0$. To all
these one must also add relation (38).

In the case of (B) class solutions, we have six arbitrary real
constants: $a$ and two constants determined by the function $g$ and
three constants given by $\zeta$.  These mean that we have six
linearly independent Killing vector fields for class (B) which will
be presented together with their Lie algebra in the next subsection.
In the case of (C), which includes (A) and (B) solutions as special
subcases, we have only one additional symmetry generator given by
(32). Although some other subcases of (C) can be defined, this case
will not be pursued any further as it has many additional
conditions.

\subsection{Killing Vector Fields and Lie Algebra for the case (B)}

In terms of the two nonzero constants $m$ and $m_0$ defined by (24),
we have obtained, in addition to three $so(3)$ 1-forms given by (5),
three additional linearly independent KY 1-forms:
\begin{eqnarray}
\omega_1&=&\cos\theta \omega-\frac{1}{m_{0}}WH_0\sin\theta
e^{2}\;,\nonumber\\
 \omega_2&=&\sin\theta\cos\varphi
\omega+\frac{1}{m_{0}}WH_0(\cos\theta\cos\varphi e^{2}-\sin\varphi e^{3})\;,\\
\omega_3&=&-\sin\theta\sin\varphi
\omega-\frac{1}{m_{0}}WH_0(\cos\theta\sin\varphi e^{2}+\cos\varphi
e^{3})\;,\nonumber
\end{eqnarray}
where the 1-form $\omega$ is defined by
\begin{eqnarray}
\omega=T^{1/m}H_2 e^{0}+WLe^{1}\;.
\end{eqnarray}
By noting that the metric dual of $e^{0}$ is
$-H_{0}^{-1}\partial_t$, in terms of $K=T^{1/m}$ the vector field
$X$ which is the metric dual of $\omega$ can be written as
\begin{eqnarray}
X=\tilde{\omega}=-K\frac{H_2}{H_0}\partial_t+\dot{K}\frac{H_2}{H^{\prime}_{0}}\partial_r\;.
\end{eqnarray}
The Killing vector fields corresponding to (39) are then given by
\begin{eqnarray}
X_1&=&\cos\theta X-MZ_1\;,\nonumber\\
X_2&=&\sin\theta\cos\varphi X+MZ_2\;,\\
X_3&=&-\sin\theta\sin\varphi X -MZ_3\;,\nonumber
\end{eqnarray}
where $M=\dot{K}H_0/m_0 H_2$ and the vector fields $Z_i$ are defined
as
\begin{eqnarray}
Z_1=\sin\theta \partial_\theta\;,\quad Z_2=\cos\theta\cos\varphi
\partial_\theta-\frac{\sin\varphi}{\sin\theta}\partial_\varphi\;,\quad
Z_3=\cos\theta\sin\varphi
\partial_\theta+\frac{\cos\varphi}{\sin\theta}\partial_\varphi\;.
\end{eqnarray}

By making use of
\begin{eqnarray}
\;[Z_1,Z_2]&=&K_2\;,\quad\quad[Z_1,Z_3]=-K_1\;,\quad
[Z_2,Z_3]=-K_3\;,\\
\;[Z_1,K_1]&=&-Z_3\;,\quad [Z_1,K_2]=Z_2\;,\quad\quad
[Z_1,K_3]=0\;,\nonumber\\
\;[Z_2,K_1]&=&0\;,\quad\quad\;[Z_2,K_2]=-Z_1\;,\;\quad [Z_2,K_3]=-Z_3\;,\\
\;[Z_3,K_1]&=&Z_1\;,\quad\;\;[Z_3,K_2]=0\;,\quad\;\;\quad
[Z_3,K_3]=Z_2\;,\nonumber
\end{eqnarray}
we obtain
\begin{eqnarray}
\;[K_1,X_1]&=& X_3\;,\quad\quad  [K_1,X_2]=0\;,\quad\quad \quad
\quad
[K_1,X_3]=-X_1\;,\nonumber\\
\;[K_2,X_1]&=&X_2 \;,\quad \quad [K_2,X_2]=-X_1\;,\quad\quad\;\;
[K_2,X_3]=0\;,\\
\;[K_3,X_1]&=& 0\;,\quad\quad\;\;  [K_3,X_2]=-X_3\;,\quad\;\;\quad
[K_3,X_3]=X_2\;.\nonumber
\end{eqnarray}
On the other hand in terms of $s=X(M)+M^{2}$ we have
\begin{eqnarray}
\;[X_1,X_2]&=& sK_2\;,\quad  [X_1,X_3]=-sK_1\;,\quad
[X_2,X_3]=-sK_3\;.
\end{eqnarray}

For evaluation of $s$,  we should recall that $K$ obeys the equation
(36). For $m=1$ we have $K=T$ and $T\ddot{T}=-m_0$, which can be
integrated to $\dot{T}^{2}+2m_0\ln T=\ell_0$. For $m\neq 1$, it can
also be easily integrated once to obtain
\begin{eqnarray}
\dot{K}^{2}=-\frac{m_0}{1-m} K^{2(1-m)}+m_2\;,
\end{eqnarray}
where $\ell_0$ and $m_2$ are integration constants determined by the
metric. We can therefore write
\begin{eqnarray}
s&=&-K\frac{H_2}{H_0}M_t+\dot{K}\frac{H_2}{H^{\prime}_{0}}M_r+M^2\;,\nonumber\\
&=&\frac{1}{m_0}[-K\ddot{K}+(1-m)\dot{K}^{2}]\;,\\
&=& \left\{
\begin{array}{cc}
\frac{1-m}{m_0}m_2\;, & \quad {\rm for}\quad m\neq 1, \\
1\;, &\quad {\rm for}\quad m=1\;,\nonumber
\end{array}
\right.
\end{eqnarray}
where  we have made use of (38) in the second line. When $m\neq 1$
and $s\neq0$ the $X_i$ generator can be normalized with the same
constant such that at the right hand side of (47), there appear $\pm
K_j$. Such a normalization does not affect relations (46). But when
the integration constant $m_{2}$ is zero, the generator set
$\{X_1,X_2,X_3\}$ form an abelian subalgebra. In that case the
symmetry algebra is isomorphic to $3D$ Euclidean algebra $e(3)$.
This shows how the symmetry of the metric may change for different
values of an integration constant.

\section{Maximal Symmetries : $\dot{T}=0,\;H_0^{\prime}\neq 0$ Solutions}

Let us begin this case by considering $\alpha$ and $\beta$ as given
in equation (10), such that $A=(H_0^{\prime}/H_1)\xi$ and the
functions $V,\;V_i,\;u$ and $v_i$ depend only on $\tau=t/T$ for
$\beta$ is independent of $r$. Since $\dot{T}=0$ in this case,
condition (13) requires that $V=0$ and we can then write, from the
first equation of (4):
\begin{eqnarray}
\alpha=U+\frac{H_0^{\prime}}{H_1}\xi\;,\quad \beta=\xi_\tau\;,
\end{eqnarray}
where $\xi$ is given by (12) and $U_\tau=0$. The third equation of
(4) now provides us with
\begin{eqnarray}
U=c_{0}H_0\;,\quad \xi_{\tau\tau}+k_1\xi=0\;,
\end{eqnarray}
where $c_0$ is an integration constant, and the metric constant
$k_{1}$ is defined by
\begin{eqnarray}
k_1=\frac{H_0^{2}}{H_1}(\frac{H_0^{\prime}}{H_0H_1})^{\prime}\;.
\end{eqnarray}
Note that the second equation of (51) is equivalent to three similar
equations for $u,v_1$ and $v_2$, each of which gives two linearly
independent solutions depending on the value of $k_1$.

Having completely specified $\alpha$ and $\beta$ with
$A=H_0^{\prime}\xi/H_1$ and $B=\xi_\tau$, we now turn to conditions
(15) and (16), which in this case amount to
\begin{eqnarray}
H_0 H_0^{\prime}=k_1H_2H_2^{\prime}\;,\quad
P^{\prime}=-\frac{H_1}{H_2^{2}}.
\end{eqnarray}
$k_1$ is a nonzero metric constant for $H_0^{\prime}\neq 0$. Both of
these relations can be integrated to
\begin{eqnarray}
H_0^{2}=k_1H_2^{2}+k_2\;,\quad P^{2}=k_3+\frac{1}{H_2^{2}}\;,
\end{eqnarray}
where $k_2$ and $k_3$ are integration constants. The first relation
of (54) implies that for $H_0^{\prime}\neq 0,\;k_1$ must be
different from zero.  The conditions (52) and (53) also imply that
\begin{eqnarray}
P^{2}=\frac{H_0^{2}}{k_2H_2^{2}}\;,\quad k_1=k_2k_3\;,
\end{eqnarray}
which means for $H_2^{\prime}\neq 0,\;k_2$ and $k_3$ must be
different from zero as well.

If $k_1$ is a negative constant such that $k_1=-\kappa^{2}$ where
$\kappa$ is a nonzero real number, we can write, in terms of
integration constants $c_i$, the $\xi$ solutions of (51) as
\begin{eqnarray}
\xi &=&(c_1\cosh\kappa\tau+c_2\sinh\kappa\tau)\cos\theta+\sin\theta\sigma^{A}\;,\nonumber\\
\sigma^{A}&=&(c_3\cosh\kappa\tau+c_4\sinh\kappa\tau)\cos\varphi-
(c_5\cosh\kappa\tau+c_6\sinh\kappa\tau)\sin\varphi\;.\nonumber
\end{eqnarray}
Then $\gamma$ and $\delta$ can be determined from (14) and (17) as
\begin{eqnarray}
\gamma=\frac{H_2^{\prime}}{H_1}\xi_{\tau\theta}+TH_2 g\;,\quad
\delta=\frac{H_2^{\prime}}{H_1}\sigma^{A}_{\tau\varphi}+TH_2(a\sin\theta+
\cos\theta g_\varphi)\;.
\end{eqnarray}

The solutions (50) and (56) determine seven linearly independent KY
$1$-forms in addition to $so(3)$ solutions. The first one is
$\omega_0=H_0e^{0}$ which corresponds to $K_0=-\partial/\partial t$.
For $k_1=-\kappa^{2}$, we can write these additional forms as
follows:
\begin{eqnarray}
\omega_{1} &=& \cos\theta\psi_1-\kappa
\frac{H_2^{\prime}}{H_1}\sinh\kappa\tau \sin\theta
e^{2}\;,\quad\;\;\;\omega_{2}=\cos\theta\psi_2-\kappa
\frac{H_2^{\prime}}{H_1}\cosh\kappa\tau \sin\theta e^{2}\;,\nonumber\\
\omega_{3} &=& \sin\theta\cos\varphi\psi_1+\kappa
\frac{H_2^{\prime}}{H_1}\sinh\kappa\tau \phi_{1}
\;,\;\;\quad\omega_{4}= \sin\theta\cos\varphi\psi_2+\kappa
\frac{H_2^{\prime}}{H_1}\cosh\kappa\tau \phi_{1}
\;,\\
\omega_{5} &=&-\sin\theta\sin\varphi\psi_1-\kappa
\frac{H_2^{\prime}}{H_1}\sinh\kappa\tau \phi_{2}\;,\quad \omega_{6}
=-\sin\theta\sin\varphi\psi_2-\kappa
\frac{H_2^{\prime}}{H_1}\cosh\kappa\tau \phi_{2} \;,\nonumber
\end{eqnarray}
where the $1$-forms $\psi_i$ and $\phi_i,\;i=1,2$ are defined by
\begin{eqnarray}
\psi_{1} &=& \cosh\kappa\tau \frac{H_0^{\prime}}{H_1} e^{0}+\kappa
\sinh\kappa\tau e^{1}\;, \quad \phi_1=\cos\theta\cos\varphi
e^{2}-\sin\varphi e^{3}\;, \nonumber\\
\psi_{2}&=&\sinh\kappa\tau \frac{H_0^{\prime}}{H_1} e^{0}+\kappa
\cosh\kappa\tau e^{1}\;,\quad \phi_2=\cos\theta\sin\varphi
e^{2}+\cos\varphi e^{3}\;.\nonumber
\end{eqnarray}
The vector fields $W_i=\tilde{\psi}_{i}$ and
$\bar{Z}_{i+1}=\tilde{\phi}_i$ are, in terms of $Z_2$ and $Z_3$
given by (43), as follows:
\begin{eqnarray}
W_1&=&-\cosh\kappa\tau\frac{H_0^{\prime}}{TH_0H_1}
 \partial_\tau+\frac{\kappa}{TH_1}
\sinh\kappa\tau \partial_r\;,\quad \bar{Z}_2=\frac{1}{TH_2}Z_2\;,\\
W_2&=&-\sinh\kappa\tau\frac{H_0^{\prime}}{TH_0H_1}
\partial_\tau+\frac{\kappa}{TH_1}\cosh\kappa\tau \partial_r\;,\quad
\bar{Z}_3=\frac{1}{TH_2}Z_3\;.
\end{eqnarray}

It is also not difficult to verify that for $H_2=r$ we have, from
(54)
\begin{eqnarray}
H_0^{2}=k_1 r^{2}+k_2\;,\quad
H_1^{2}=\frac{1}{1+k_3r^{2}}\;,\nonumber
\end{eqnarray}
and $H_0H_1=1$ for $k_2=1$. In the case of $k_1=-1=k_3$ and $k_2=1$,
we recover the static form of the de Sitter metric (see
\cite{Thirring} pp.492). KY 1-forms for five different forms of the
de Sitter type space-times are obtained in the next section.

\section{$\rm{de}$ Sitter and Robertson-Walker Type Symmetries: $H_0^{\prime}=0,\dot{T}\neq 0$}

Since $\alpha_t=0$ in this case, it is convenient to start by
defining the constant
\begin{eqnarray}
\ell=T^{2}\partial_t\frac{\dot{T}}{T}\;.
\end{eqnarray}
The nonzero and zero values of $\ell$ will then be considered
separately. In these two cases, $T$ is restricted to be a special
function of time by the symmetry requirement.

Case A considered below leads us to a family of de Sitter type
space-times with ten independent Killing vector fields and,
depending on the values of $\ell$ and other integration constants,
four different time evolution regimes can explicitly be specified by
the symmetry requirement. The B case, specified by $\ell=0$,
corresponds to the best known form of de Sitter space-time, again
having ten independent Killing vector fields such that $T$ is an
exponential function of time. However, there is an important special
case specified by $\alpha=0$, and therefore $T$ is not restricted by
any symmetry requirement. This corresponds to the Robertson-Walker
space-time with six dimensional symmetry algebra in which they are
the Einstein equations that give the time dependence as shown in
Table I.

\subsection{The case $\ell\neq 0$}

We start with
\begin{eqnarray}
\alpha=U(r)+H_2\zeta\;, \quad \beta=\dot{T}Y(r)+B\;,
\end{eqnarray}
where $\zeta$ is given by (33) and $B$ is defined as in Section III.
Condition (13) implies that $U=H_0PY$ and the second and third
equations of (4) then yield
\begin{eqnarray}
Y^{\prime}&=&H_1PY\;,\quad\quad\quad\; Y=-\frac{H^{2}_0}{\ell H_{1}}(PY)^{\prime}\;,\\
B_{r}&=&\dot{T}\frac{H_1H_2}{H_0}\zeta\;,\;\;
T^{2}\partial_t\frac{B}{T}=-H_0H_2P\zeta\;.
\end{eqnarray}
These are the reduced forms of equations (20) and (21). The first
equation of (62) gives $Y=c_1H_2$ and from the second equation we
then obtain
\begin{eqnarray}
P^{\prime}=-H_1(P^{2}+\frac{\ell}{H_0^{2}})\;.
\end{eqnarray}
The two equations of (63) imply that, in terms of
\begin{eqnarray}
\eta_2=b_1\cos\theta+\sin\theta
\sigma^{b}\;,\;\;\sigma^{b}=b_2\cos\varphi-b_3\sin\varphi\;,
\end{eqnarray}
the most general solution for $B$ is of the form
$B=\dot{T}\eta_1(r,\theta,\varphi)+T\eta_2(\theta,\varphi)$. The
second equation of (63) specifies $\eta_1$ in terms of $\zeta$:
\begin{eqnarray}
\eta_1=-\frac{H_0}{\ell}H_2P\zeta\;,
\end{eqnarray}
and the first equation yields nothing but condition (64).

Having completely specified $\alpha$ and $\beta$ with $A=H_2\zeta$
and
\begin{eqnarray}
B=-\frac{H_0}{\ell}\dot{T}H_2P\zeta+T\eta_2=\sin\theta\sigma^{B}+
\cos\theta(b_1T-\frac{H_0}{\ell}\dot{T}H_2P\sigma)\;,
\end{eqnarray}
such that
\begin{eqnarray}
\sigma^{B}=T\sigma^{b}-\frac{1}{\ell}\dot{T}H_0H_2P\sigma\;,
\end{eqnarray}
the condition (14) amounts to
\begin{eqnarray}
P^{2}=-\frac{\ell}{H_0^{2}}+\frac{1}{H_2^{2}}\;.
\end{eqnarray}
It is not difficult to verify that condition (64) is implied by
(69), which also yields $P^{\prime}=-H_1/H_2^{2}$. (Note that the
integration of (64) results in an additional constant multiplying
the second term of (69)). In view of this relation, condition (16)
is identically satisfied. That is, we only have condition (69).

We can now turn to (14) and (17) to evaluate $\gamma$ and $\delta$,
and the solutions can be collated as follows:
\begin{eqnarray}
\alpha&=& c_1H_0H_2P+H_2\zeta\;,\nonumber\\
\beta &=& c_1\dot{T}H_2+\sin\theta\sigma^{B}+
\cos\theta(b_1T-\frac{H_0}{\ell}\dot{T}H_2P\sigma)\;,\nonumber\\
\gamma &=&-\dot{T}\frac{H_2^{2}}{H_0}\zeta_\theta+
\dot{T}H_2P[\cos\theta\sigma^{B}-\sin\theta(b_1T-\frac{H_0}{\ell}\dot{T}H_2P\sigma)]+TH_2 g\;,\\
\delta
&=&-\dot{T}\frac{H_2}{H_0}\sigma_\varphi+H_2P\sigma^{B}_\varphi+TH_2(a\sin\theta+\cos\theta
g_\varphi) \;.\nonumber
\end{eqnarray}
Note that for $H_0=1$ and $H_{2}=r$, condition (69) yields
\begin{eqnarray}
H^2_1=\frac{1}{1-\ell r^{2}}\;.
\end{eqnarray}
These are the characteristics of the de Sitter and Robertson-Walker
space-times. It should be emphasized that the right hand side of
condition (69) must be positive, which reflects the fact that
$1/r^{2}>\ell$ is required to avoid any singularity in the
corresponding space-times.

We now turn to the explicit evolution of $T$.  By multiplying both
sides of equation (60) by $\dot{T}/T^{3}$, it can be integrated to
\begin{eqnarray}
\dot{T}=\epsilon(\ell_3T^{2}-\ell)^{1/2}\;,
\end{eqnarray}
and then by integrating once more we get
\begin{eqnarray}
T=\left\{
\begin{array}{ll}
(\frac{\ell}{\ell_3})^{1/2}\cosh(\epsilon \ell_3^{1/2}t+a)\;;&\;\;{\rm for}\;\ell_3>0\;,\;\ell>0\;, \\
\frac{\ell_{0}}{\ell_3^{1/2}}\sinh(\epsilon\ell_3^{1/2}t+a)\;;&
\;\;{\rm for}
\;\ell_3>0\;,\;\ell=-\ell_{0}^{2}<0\;,\\
\frac{\ell_{0}}{k_0}\sin(\epsilon k_{0} t+a)\;; &\;\; {\rm for} \;
\ell_3=-k_{0}^{2}<0\;,\;\ell=-\ell_{0}^{2}<0\;,\\
\epsilon \ell_{0} t+a\;; &\;\; {\rm for}\;
\ell_3=0,\;\ell=-\ell_{0}^{2}<0\;,
\end{array}
\right.
\end{eqnarray}
where $\ell_3$ and $a$ are integration constants and
$\epsilon=\pm1$. Note that the third solution can be inferred from
the second one and that all the corresponding KY 1-forms can
explicitly be read from (70). All of these de Sitter space-times of
which the fourth one is a flat space-time, are spaces of constant
curvature with the curvature scalar given by $\ell_3$ (see Appendix
B).

\subsection{de Sitter type Symmetries: $\ell=0=H_0^{\prime}$}

The condition $\ell=0$ is equivalent to $\dot{T}=\lambda T$, that is
to
\begin{eqnarray}
T=T_0\exp(\lambda t) \;,\nonumber
\end{eqnarray}
where $\lambda$ is a metric constant and $T_0$ is an integration
constant. In this case, to avoid excessive repetitions, we shall be
content to present the solutions:
\begin{eqnarray}
\alpha&=& c_0H_0+\frac{1}{H_0P}\zeta\;,\nonumber \\
\beta&=&\dot{T}[\frac{c_0}{P}+\frac{1}{2}(\frac{1}{\dot{T}^{2}}+\frac{1}{P^{2}})\zeta+
b_1\cos\theta+\sin\theta \sigma^{b}]\;,\nonumber\\
\gamma
&=&\dot{T}H_2P\{[-\frac{1}{H_0^{2}P^{2}}+
\frac{1}{2}(\frac{1}{\dot{T}^{2}}+\frac{1}{P^{2}})]\zeta_{\theta}
-b_1\sin\theta+\cos\theta \sigma^{b}\}+TH_2 g\;,\\
\delta &=&\dot{T}H_2P\{[-\frac{1}{H_0^{2}P^{2}}+
\frac{1}{2}(\frac{1}{\dot{T}^{2}}+\frac{1}{P^{2}})]\sigma_{\varphi}
+\sigma_{\varphi}^{b}\}+TH_2(a\sin\theta+\cos\theta g_\varphi)
\;,\nonumber
\end{eqnarray}
which provide us with ten-dimensional symmetry algebra. Here,
$\sigma^{b}$ is given by (65). The above solutions can be easily
verified provided that the condition $H_2^{\prime}=\epsilon H_1$
holds, which implies that $H_2P=\epsilon$ and hence,
$P^{\prime}=-H_1P^{2}$.

\subsection{The Case $\alpha=0$:
Robertson-Walker Symmetries}

When $\alpha=0$, the second and third equations of (4) imply that
$\beta_r=0$ and
\begin{eqnarray}
\beta=B=T(\sin\theta \sigma+a_3\cos\theta)\;,
\end{eqnarray}
with $\sigma=\sigma^{B}/T=a_1\cos\varphi-a_2\sin\varphi$. In that
case, eq. (15) is identically satisfied and eq. (16) gives
$P^{\prime}=-H_1/H^{2}_2$, which can be integrated to yield the
second relation of eq. (54) and provide us with
\begin{eqnarray}
H^2_1=\frac{1}{1+k_3 r^{2}}\;,
\end{eqnarray}
for $H_0=1$ and $H_{2}=r$. Then the solutions can be easily read
from (14) and (17) to be
\begin{eqnarray}
\gamma &=& T\frac{H^{\prime}_{2}}{H_1}(\cos\theta \sigma-a_3\sin\theta)+TH_2 g\;,\nonumber\\
\delta
&=&T\frac{H^{\prime}_2}{H_1}\sigma_\varphi+TH_2(a\sin\theta+\cos\theta
g_\varphi)\;.\nonumber
\end{eqnarray}
With $\alpha=0$ and $\beta$ as in (75), the following Killing vector
fields are obtained :
\begin{eqnarray}
I_1 &=& \frac{1}{H_1}\sin\theta\cos \varphi\partial_r+PZ_2
\;,\nonumber\\
I_2 &=&-\frac{1}{H_1}\sin\theta\sin \varphi\partial_r-PZ_3
\;,\\
I_3 &=& \frac{1}{H_1}\cos \theta\partial_r-PZ_1\;.\nonumber
\end{eqnarray}
These are the generalized translation generators which, in the usual
cartesian coordinates read, respectively, as $H_1^{-1}\partial_x\;,
-H_1^{-1}\partial_y$ and $H_1^{-1}\partial_z$. Together with the
$so(3)$ generators, they close into the following Lie algebra
structure:
\begin{eqnarray}
\;[I_1,I_2] &=& k_3 K_3\;,\quad\;\quad [I_2,I_3]=k_3 K_1\;,\quad
[I_1,I_3]=-k_3 K_2\;,\nonumber\\
\;[K_1,I_2] &=& -I_3=[K_2,I_1]\;,\;\;[K_1,I_3]=I_2=-[K_3,I_1]\;,\\
\;[K_2,I_3]&=&I_1=[K_3,I_2]\;,\quad\;\; [K_i,I_i]= 0\;,\quad
i=1,2,3\;.\nonumber
\end{eqnarray}

\section{Flat Space-Time Solutions: $\dot{T}=0,\;H_0^{\prime}=0$}

In this case, $\alpha$ is independent of $\tau=t/T$, $\beta$ is
independent of $r$, and condition (13) requires that $V=0$.
Therefore it is convenient to start with $\alpha=c_0H_0+\eta_1$ and
$\beta=\eta_2$, where $c_0$ is a constant and $\eta_i$ are as in
(65) of the previous section, such that $\eta_1$ is independent from
$\tau$ and $\eta_2$ is independent from $r$.  The third equation of
(4) gives $\eta_{2\tau}=-H_0\eta_{1r}/H_1$. Since the left hand side
of this equality depends on $\tau$ and is independent of $r$, but
the right hand side depends on $r$ and is independent of $\tau$,
both sides must be equal
\begin{eqnarray}
\eta_{2\tau}=\zeta=-\frac{H_0}{H_1}\eta_{1r} \;,
\end{eqnarray}
where $\zeta$ is given by (33). From the first equality, we obtain
$\eta_2=\tau\zeta+\zeta^{(b)}$ where $\zeta^{(b)}$ is the same as
$\zeta$ with $a=(a_1,a_2,a_3)$ replaced by $b=(b_1,b_2,b_3)$. On the
other hand, for $A=\eta_1$ and $B=\tau\zeta+\zeta^{(b)}$, conditions
(15) and (16) respectively yield
\begin{eqnarray}
\eta_{1\theta}=-\frac{H_2^{2}P}{H_0}\zeta_{\theta}\;,\quad
P^{\prime}=-\frac{H_1}{H_2^{2}}\;.
\end{eqnarray}
In view of the second equality of (79), we obtain
$(H_2^{2}P)^{\prime}=H_1$ from the integrability condition
$\eta_{1\theta r}=\eta_{1r\theta}$. From the second equation of (79)
and $(H_2^{2}P)^{\prime}=H_1$, we get $P^{2}=1/H_2^{2}$, which can
also be obtained from the integration of $P^{\prime}=-H_1/H_2^{2}$
with the zero integration constant (see the discussion following
equations (53) and (54)).

From (14) and (17), one can easily write out $\gamma$ and $\delta$
together of $\alpha$ and $\beta$ to be as :
\begin{eqnarray}
\alpha&=& c_0H_0 -\frac{1}{H_0P}\zeta\;,\quad
\beta=\tau\zeta+\zeta^{(b)}\;,\nonumber\\
\gamma &=&H_2 P(\tau\zeta+\zeta^{(b)})+TH_2 g\;,\\
\delta
&=&H_2P(\tau\sigma_{\varphi}+\sigma^{(b)}_{\varphi})+TH_2(a\sin\theta+\cos\theta
g_\varphi) \;.\nonumber
\end{eqnarray}
We have ten linearly independent KY 1-forms and only one condition
$P^{2}=1/H_2^{2}$ which is equivalent to $H_2^{\prime 2}=H_1^{2}$.
For $H_2=r$, we have $H_1=\pm1$, and the corresponding Killing
vector fields are directly determined from (81) with $P=\pm1/r$.

\begin{acknowledgments}
This work was supported in part by the Scientific and Technical Research Council of
Turkey (T\"{U}B\.{I}TAK).
\end{acknowledgments}

\appendix

\section{Metric Constants}

In Section IV, the following four metric constants were defined :
\begin{eqnarray}
m&=&L^{\prime}\frac{H_0}{H_1H_2}\;,\;\;
m_0=\frac{H^{2}_0}{LH_1}(\frac{H_2}{H_{0}})^{\prime}\;,\\
k&=&\frac{H^{\prime}_0H_2}{H_0H_2^{\prime}}\;,\;\;\quad
k_0=-\frac{H_0^{2}}{H_1H_2}(\frac{H_2^{\prime}}{H_1})^{\prime}\;.
\end{eqnarray}
The first relation of (A1) can be integrated to find
\begin{eqnarray}
H^{\prime}_0=\ell_1H^{-m}_{0}H_1 H_2\;,\quad
L=\frac{1}{\ell_1}H_{0}^{m}\;,
\end{eqnarray}
which is also valid for $m=0$. The second relation of (A1) can be
arranged as
\begin{eqnarray}
\frac{H_{2}^{\prime}}{H_2}=\frac{H^{\prime}_{0}}{H_{0}}[m_{0}
(\frac{H_{1}}{H^{\prime}_{0}})^{2}+1] \;.
\end{eqnarray}
The first relation of (A2) can also be integrated to find
$H_0=\ell_2H_2^{k}$, where $\ell_1$ and $\ell_2$ are also non-zero
metric constants. By combining (A4) with the first relation of (A2),
we get
\begin{eqnarray}
(1-k)H^{\prime\;2}_0=m_0kH^{2}_{1}\;,
\end{eqnarray}
and then, by (A3) and $ H_0=\ell_2H_2^{k}$, we arrive at
\begin{eqnarray}
(1-k)=m_0k(\frac{\ell_2^{m}}{\ell_1})^{2}H^{2(mk-1)}_{2}\;.
\end{eqnarray}
Thus for nonconstant $H_2$, we must have
\begin{eqnarray}
km=1\;,\quad m=1+m_0(\frac{\ell_2^{m}}{\ell_1})^{2}\;.
\end{eqnarray}
Note that $k=1$ if and only if $m_0=0$ and if and only if $m=1$ for
$H_{2}^{\prime}\neq 0$.

\section{Curvature Forms}

The curvature 2-forms
$R_{ab}=d\omega_{ab}+\omega_{ac}\wedge\omega^{c}_{\;b}$ for the
considered class of spherically symmetric space-times are computed
by using eqs. (2) and (3) to be as follows :
\begin{eqnarray}
R_{01}&=&(S+\frac{h_{0}^2+
h_0^{\prime}-h_0h_1}{T^2H_{1}^2})e^{01}\;,\;\;R_{12}=S_2e^{02}+S_1e^{12}\;,\nonumber\\
R_{02}&=&(S+\frac{h_0 h_2}{T^2
H_{1}^2})e^{02}+S_2e^{12}\;,\quad R_{13}=S_2e^{03}+S_1e^{13}\;,\\
R_{03}&=&(S+\frac{h_0 h_2}{T^2 H_{1}^2})e^{03}+S_2e^{13}\;,\quad
R_{23}=(S_1+\frac{H_{1}P^{\prime}+(H_1/H_2)^{2}}{T^2H_{1}^2})e^{23}\;,\nonumber
\end{eqnarray}
where
\begin{eqnarray}
S=-\frac{\ddot{T}}{T H_{0}^2}\;,\quad S_1=\frac{\dot{T}^2}{T^2
H_0^2}-\frac{P^{\prime}+H_1P^{2}}{T^2 H_1}\;,\quad
S_2=\frac{\dot{T}}{T^2}\frac{h_0}{H_0 H_1}\;.
\end{eqnarray}
Note that $S=0=S_2$ when $\dot{T}=0$. The corresponding Ricci
1-forms and curvature scalar can be found from $P_b=i_{X^a}R_{ab}$
and $\Re=i_{X^b}P_b$.

For $\dot{T}=0=H_0'$ and $H_2'^2=H_1^2$ we have $S_2=0$ in addition
to $S=0=S_1$. It then follows that all the curvature components
vanish, and we obtain the Minkowski space-time. In the case of the
static form of de Sitter space-time, that is for $\dot{T}=0,\;H_2=r$
and
\begin{eqnarray}
H_0^{2}=k_1 r^2+k_2\;,\; H_1^{2}=\frac{1}{1+k_3
r^2}\;,\;k_1=k_2k_3\nonumber
\end{eqnarray}
we have $S=0=S_2$ and $S_1=-k_3/T^2$ which gives the constant
curvature solutions
\begin{eqnarray}
R_{0j}=\frac{k_3}{T^2}e^{0j}\;,\quad
R_{ij}=-\frac{k_3}{T^2}e^{ij}\;,
\end{eqnarray}
where $i,j=1,2,3$ and $i<j$. These two space-times are maximally
symmetric with constant curvature.

For $\dot{T}=\lambda T\;,H_0'=0$ and $H_2'=\epsilon H_1$ which
indicate de Sitter space-time, we have
\begin{eqnarray}
S=-\frac{\lambda^{2}}{H_0^2}=-S_1\;, \quad S_2=0\;.\nonumber
\end{eqnarray}
These lead us again to the curvature solution given by (B3) provided
that $k_3/T^2$ is replaced by $-\lambda^{2}/H_0$. For the
Robertson-Walker space-time, we have
\begin{eqnarray}
R_{0j}=-\frac{\ddot{T}}{T}e^{0j}\;,\quad
R_{ij}=[(\frac{\dot{T}}{T})^{2}+\frac{k}{T^2}]e^{ij}\;.
\end{eqnarray}

Finally, for the four forms of de Sitter space-time found in Section
VI, we have
\begin{eqnarray}
H_0=1\;,\quad H^{2}_1=\frac{1}{1-\ell r^{2}} \quad H_2=r\;,\nonumber
\end{eqnarray}
and $T$ expressions are explicitly given by equations (73). For
these values we find $S_2=0$ and $S_1=\ell_3=-S$, where $\ell_3$ is
a constant. Thus
\begin{eqnarray}
R_{0j}=-\ell_3e^{0j}\;,\quad R_{ij}=\ell_3e^{ij}\;.
\end{eqnarray}
where we have used $\dot{T}=\epsilon(\ell_3T^{2}-\ell)^{1/2}$ found
in equation (72).


\newpage

\begin{table}[tbp]
\caption{ Metric coefficient functions and the numbers of linearly
independent KY-forms of some well-known spherically symmetric
space-times. For all these cases $H_2$ is the radial coordinate $r$.
The numbers of the fifth column represent the dimensions $d(4,1)$ of
the corresponding symmetry algebras whose common part consists of
three $so(3)$ generators given by the equation (6) of the main text.
The last two columns denote the numbers $d(4,2)$ and $d(4,3)$ of the
linearly independent KY 2-forms and 3-forms which are explicitly
calculated in the accompanying paper. The fifth rows represent three
different forms of the de Sitter space-time, of which the third
consists of four cases with different time evolutions given in
Section VI. The numbers $d(n,p)$ for the maximally symmetric
space-times, such as those of Minkowski
and de Sitter, represent the upper bounds for dimension $n=4$.} 

\bigskip

\begin{tabular}{|c|c|c|c|c|c|c|c|}
\tableline\tableline
Space-time &\quad $T$ \quad & \quad $H_{0}$\quad & \quad $H_{1}$ \quad & $d(4,1)$\quad
& $ d(4,2)$\quad & $d(4,3)$\quad\\
\tableline\tableline
\quad {\bf Schwarzschild} \quad &\quad $1$ \quad &\quad
$\sqrt{1-\frac{2M}{r}}$ \quad
&\quad $\frac{1}{\sqrt{1-\frac{2M}{r}}}$ \quad & 4 & 1 & 0\\
\tableline
\quad {\bf Reissner-Nordstr{\o}m} \quad &\quad $1$ \quad &\quad
$\sqrt{1-\frac{2M}{r}+\frac{Q^{2}}{r^{2}}}$ \quad &\quad
$\frac{1}{\sqrt{1-\frac{2M}{r}+\frac{Q^{2}}{r^{2}}}}$ \quad & 4 & 1 & 0\\
\tableline \quad {\bf Robertson-Walker}
\quad & $T(t)$ & $1$ & $\frac{1}{\sqrt{1-kr^{2}}}$ & 6 & 4 & 1 \\
\tableline
\quad  \underline{{\bf de Sitter}} \quad &  &  &  & & & \\
\quad static form \quad &\quad $1$ \quad &\quad $\sqrt{1-kr^{2}}$
\quad &\quad
$\frac{1}{\sqrt{1-kr^{2}}}$ \quad  & 10 & 10 & 5 \\
\quad usual form
 & $e^{(\Lambda/3)^{1/2}t}$ & 1 & 1 & 10 & 10 & 5\\
\quad four additional forms \quad &\quad
$\dot{T}^{2}=\ell_3T^{2}-\ell$ \quad &\quad $1$ \quad &\quad
$\frac{1}{\sqrt{1-\ell\;r^{2}}}$ \quad  &
10 & 10 & 5\\
& & & & & &\\
\tableline
\quad {\bf Minkowski} \quad  &\quad $1$ \quad & \quad $1$ \quad
&\quad $1$ \quad & 10 & 10 & 5 \\
\tableline \tableline
\end{tabular}
\end{table}

\end{document}